\documentclass[prd,twocolumn,tightenlines,showpacs,nofootinbib,superscriptaddress,nobalancelastpage]{revtex4-1}

\usepackage{graphics}
\usepackage{graphicx}
\usepackage{longtable}
\usepackage{url}
\usepackage{bm}
\usepackage{natbib}
\usepackage{textpos}
\usepackage{amsfonts,amsmath,amssymb,mathrsfs,bm,amsthm}
\usepackage{float}
\restylefloat{table}
\usepackage{booktabs}
\usepackage{array}
\usepackage{tabu}
\usepackage{dcolumn}
\usepackage{rotating}
\usepackage{ulem}

\usepackage{amsmath}
\usepackage{amssymb}
\usepackage{graphicx}
\usepackage{epstopdf}
\usepackage{inputenc}

\usepackage{euscript}

\newcommand{\RN}[1]{%
  \textup{\uppercase\expandafter{\romannumeral#1}}%
}

\usepackage{nicefrac}
\usepackage{amsthm}
\usepackage{color}
\usepackage{cancel}
\usepackage{appendix}
\usepackage{makecell}

\usepackage[colorlinks=true,linkcolor=black,citecolor=blue,urlcolor=blue,bookmarksopen]{hyperref}
\newcommand{\appropto}{\mathrel{\vcenter{
  \offinterlineskip\halign{\hfil$##$\cr
    \propto\cr\noalign{\kern2pt}\sim\cr\noalign{\kern-2pt}}}}}
\renewcommand{\v}[1]{\boldsymbol{#1}}		%bold-math for vectors

\def\prn#1{{\left(#1\right)}}

\def\sbrk#1{{\left[#1\right]}}
\def\abrk#1{{\langle#1\rangle}}

\def\bra#1{{\langle#1|}}

\def\cg(#1,#2)(#3,#4)(#5,#6){\bra{#1,#2,#3,#4}#5,#6\rangle}

\def\ts#1{{_{\mbox{\scriptsize #1}}}}

\def\threej(#1,#2)(#3,#4)(#5,#6){\begin{pmatrix}#1&#3&#5\\#2&#4&#6\end{pmatrix}}
\def\sixj(#1,#2,#3)(#4,#5,#6){\begin{Bmatrix}#1&#2&#3\\#4&#5&#6\end{Bmatrix}}
\def\ninej(#1,#2,#3)(#4,#5,#6)(#7,#8,#9){\begin{Bmatrix}#1&#2&#3\\#4&#5&#6\\#7&#8&#9\end{Bmatrix}}

\def\mb{\mathbf}
\def\bs{\boldsymbol}

\begin{document}

%%%%%%%%%%%%%%%%%%%%%%%%

\title{Gravity Probe Spin:\\ Prospects for measuring general-relativistic precession \\ of intrinsic spin using a ferromagnetic gyroscope}

%%%%%%%%%%%%%%%%%%%%%%%%

\date{\today}

\author{Pavel~Fadeev}
\email{pavelfadeev1@gmail.com}
\affiliation{Helmholtz Institute Mainz, Johannes Gutenberg University, 55099 Mainz, Germany}

\author{Tao~Wang}
\affiliation{Department of Physics, Princeton University, Princeton, New Jersey 08544, USA}

\author{Y.~B.~Band}
\affiliation{Department of Chemistry, Department of Physics, Department of Electro-Optics, and the Ilse Katz Center for Nano-Science, Ben-Gurion University, Beer-Sheva 84105, Israel}

\author{Dmitry~Budker}
\affiliation{Helmholtz Institute Mainz, Johannes Gutenberg University, 55099 Mainz, Germany}
\affiliation{Department of Physics, University of California at Berkeley, Berkeley, California 94720-7300, USA}

\author{Peter~W.~Graham}
\affiliation{Department of Physics, Stanford Institute for Theoretical Physics, Stanford University, California 94305, USA}

\author{Alexander~O.~Sushkov}
\affiliation{Department of Physics, Boston University, Boston, Massachusetts 02215, USA}

\author{Derek~F.~Jackson~Kimball}
\email{derek.jacksonkimball@csueastbay.edu}
\affiliation{Department of Physics, California State University - East Bay, Hayward, California 94542-3084, USA}

%%%%%%%%%

\begin{abstract}
An experimental test at the intersection of quantum physics and general relativity is proposed: measurement of relativistic frame dragging and geodetic precession using intrinsic spin of electrons. The behavior of intrinsic spin in spacetime dragged and warped by a massive rotating body is an experimentally open question, hence the results of such a measurement could have important theoretical consequences. Such a measurement is possible by using mm-scale ferromagnetic gyroscopes in orbit around the Earth. Under conditions where the rotational angular momentum of a ferromagnet is sufficiently small, a ferromagnet's angular momentum is dominated by atomic electron spins and is predicted to exhibit macroscopic gyroscopic behavior. If such a ferromagnetic gyroscope is sufficiently isolated from the environment, rapid averaging of quantum uncertainty via the spin-lattice interaction enables readout of the ferromagnetic gyroscope dynamics with sufficient sensitivity to measure both the Lense-Thirring (frame dragging) and de Sitter (geodetic precession) effects due to the Earth.
\end{abstract}

%%%%%%%%%

\maketitle

%%%%%%%%%%%%%%%%%%%%%%%%%%%%%%%%%%%%%%%%%%%%%%%%%%%%%%%%%%%%%%%%%%%%%%%%%%%%%

One of the most perplexing problems in theoretical physics is devising a framework encompassing Einstein's theory of general relativity (GR) and quantum mechanics (QM) \cite{Rov07,Kie12,Eichhorn2019}. Experimentally addressing this subject likely requires probing distances at the Planck scale, far too short to be reached in the near future \cite{Peskin95}. Even at longer distances, there has been a dearth of experiments at probing regimes where both GR and QM are essential to understand observations \cite{How18,Hossenfelder2018}. While quantum systems have been used in measurements of gravitational phenomena, for most such experiments the measured phenomena are either not inherently quantum mechanical (e.g. atomic measurements of the gravitational redshift \cite{Pou60,Mul10,Cho10}, where clocks are tools to observe time dilation) or the gravitational phenomena are not inherently relativistic (e.g. observations of the quantum behavior of neutrons in Earth's gravitational field \cite{Col75,Nes02}, understandable with Newtonian gravity).

We propose an experiment testing phenomena that involve both GR and QM: measurement of gravitational frame dragging \cite{LT1918} and geodetic precession \cite{Sitter1916}, which are fundamentally general-relativistic effects, with intrinsic spin, which is a fundamentally quantum phenomenon. It is crucial to emphasize that whether or not intrinsic spins undergo general relativistic precession is an experimentally open question: to date there has been no viable way to reach the required sensitivity for direct observation of frame dragging or geodetic precession of intrinsic spins. The significance of such a test is evident from the fact that GR incorporates only classical angular momentum arising from the rotation of finite-size, massive bodies \cite{MTW,Zee,Will}. The key point is that GR explicitly describes effects related to angular momentum arising from the motion of mass-energy through spacetime, but does not explicitly consider effects related to spin, where the angular momentum arises from an intrinsic quantum property of point-like particles.

Heuristically, it can be argued based on Einstein's equivalence principle that intrinsic spin should behave in the same way as the angular momentum of a classical gyroscope \cite{Mas00,Mas13,Adl12}. Thus a reasonable theoretical approach is to use standard quantum field theory for the locally flat spacetime and treat frame dragging and geodetic precession as small perturbations to the Lorentz metric \cite{Mas00,Mas13,Adl12,Silenko2005,Silenko2007,Silenko2008}. However, whether or not this theoretical approach is correct remains to be proven experimentally \cite{Tas12}; in this sense, the proposed experiment can be envisioned as an equivalence principle test in a new regime. The proposed experiment is based on electron spins; meanwhile, frame-dragging also causes light polarization to rotate \cite{Kop02}, a measurement of which would probe the analogous effect on photon spins \cite{Ser05,Rug07}.

Indeed, without guidance from experimental measurements, there are a number of open theoretical possibilities. Even at an early stage it was realized that extending GR to include effects related to intrinsic spin (as, for example, in Cartan's theory \cite{Car22}) could change the microscopic structure of GR in fundamental ways, such as introducing torsion \cite{Sab94,Ni10}. In Einstein's GR, mass-energy generates and interacts with curvature of spacetime but the torsion is zero, and so vectors curve along geodesics via parallel transport but do not twist. In Cartan's extension, intrinsic spin generates and interacts with nonzero torsion, and so frames transported along geodesics curve due to the effect of mass-energy and twist due to the effect of intrinsic spin (see, for example, the review by Hehl {\textit{et al.}} \cite{Heh76}). Thus warping of spacetime described by GR with torsion does not affect intrinsic spin in the same way as classical angular momentum, leading to order unity differences between general-relativistic precession observed with intrinsic spin and that observed with a classical gyroscope \cite{Audretsch81}. Furthermore, spin-gravity interactions deviating from the predictions of GR are common features of theories attempting to go beyond standard physics \cite{Ori09,Cap11,Saf18}. Thus the results of an experiment measuring general-relativistic precession with intrinsic spins would have important consequences regardless of the outcome, distinguishing between a number of different theoretical possibilities.

A measurement of general-relativistic precession effects using intrinsic spin can be viewed as a ``$\mathfrak{g}-1$'' test for gravity, in analogy to the $g-2$ experiments that test quantum electrodynamics \cite{Han08}, where $g$ is the electron gyromagnetic ratio. In the proposed experiment, the parameter $\mathfrak{g}$ is the gyrogravitational ratio: the ratio between intrinsic spin and angular momentum coefficients in the theoretical description of relativistic precession. If gravity affects intrinsic spin identically to orbital angular momentum, then $\mathfrak{g}=1$, as expected based on Einstein's equivalence principle applied to intrinsic spin \cite{Adl12,Overduin2015,Singh2000,Pavlichenkov09,Ni2016}. In other approaches $\mathfrak{g}$ differs from unity: for example, $\mathfrak{g}=2$ in Refs.~\cite{Nair2018,Hehl1990} and $\mathfrak{g}=3$ in Ref.~\cite{Audretsch81}.  

Such an experiment only recently became possible, even in principle, based on a proposal for a ferromagnetic gyroscope (FG) with unprecedented sensitivity \cite{Kim16}. An ideal FG is a freely floating ferromagnet whose intrinsic spin $\mb{S}$ has far greater magnitude than any rotational angular momentum $\mb{L}$ associated with precession of the ferromagnet,
\begin{align}
S \approx N\hbar \gg L \approx I\Omega~,
\label{Eq:FG-precession-regime}
\end{align}
where $N$ is the number of polarized spins in the ferromagnet, $\hbar$ is Planck's constant, $I$ is the moment of inertia of the ferromagnet, and $\Omega$ is the precession frequency. Under these conditions, in the absence of external torques, angular momentum conservation keeps the expectation value of the total angular momentum $\abrk{\mb{J}}=\abrk{\mb{S}+\mb{L}}$ fixed with respect to the local space coordinates. The spin-lattice interaction keeps $\mb{S}$ oriented along the easy magnetic axis $\hat{\mb{n}}$ and rapidly averages components of $\mb{S}$ transverse to $\hat{\mb{n}}$. This rapid averaging of transverse spin components without inducing a random walk of $\abrk{\mb{J}}$ significantly reduces quantum noise for measurement times longer than the characteristic time scale of the spin-lattice interaction, which is $\lesssim 10^{-9}$~s in most cases. This enables exquisitely precise measurements of spin precession, as discussed in detail in Refs.~\cite{Kim16,Band2018}. A number of groups are actively working on developing the requisite experimental tools to construct an FG \cite{Tao2019,Gie19,Vin19,Tim19,Huillery2020,Vin20}, opening the possibility of observing relativistic frame dragging of $\mb{S}$ as we describe below.

Specifically, we investigate measurement of both the Lense-Thirring effect \cite{LT1918,Schiff60} (frame dragging) and the de Sitter (geodetic precession) effect \cite{Sitter1916,Schouten1918,Fokker1920}. Both effects cause precession of a gyroscope orbiting a massive body such as the Earth: Lense-Thirring precession is caused by spacetime being dragged by the rotation of a massive body whereas de Sitter precession is caused by the motion of a gyroscope through spacetime curved by a mass (present also for a non-rotating massive body). The Lense-Thirring precession is characterized by the angular velocity vector \cite{Schiff60},
\begin{align}
\v{\Omega}\ts{LT} \approx \mathfrak{g} \frac{2}{5} \frac{G M}{c^2 R} \sbrk{ 3 \prn{ \v{\Omega}_E \cdot \hat{\v{R}} } \hat{\v{R}} - \v{\Omega}_E }~,
\label{eq:LT}
\end{align}
where $\mathfrak{g}$ is the gyrogravitational ratio,
%(as discussed above, $\mathfrak{g}=1$ if intrinsic spin behaves identically to orbital angular momentum in GR \cite{Mas00,Mas13,Adl12,Silenko2005,Silenko2007,Silenko2008,Overduin2015,Singh2000,Pavlichenkov09,Ni2016})
$G$ is Newton's gravitational constant, $M$ is the mass of the Earth, $\v{R} = R\hat{\v{R}}$ is the position of the satellite relative to the center of the Earth, $c$ is speed of light, and $\v{\Omega}_E$ is Earth's angular velocity ($\Omega_E \approx 2\pi \times 11.6~{\rm \mu Hz}$). For a satellite instantaneously above the North pole at $R \approx R_E \approx 6.5 \times 10^6~{\rm m}$ (where $R_E$ is Earth's radius), $\Omega\ts{LT} \approx 4 \times 10^{-14}~{\rm s^{-1}}$ for $\mathfrak{g}=1$. The de Sitter precession in a near-Earth orbit is \cite{Schiff60,Ruffini2003}
\begin{align}
\v{\Omega}\ts{dS} \approx \mathfrak{g} \frac{3}{2} \frac{G M}{c^2R^2} \prn{ \hat{\v{R}} \times \v{v} }~,
\label{eq:dS}
\end{align}
where $\v{v}$ is the satellite velocity. For the same satellite at $R \approx R_E$ one obtains $\Omega\ts{dS} \approx 10^{-12}~{\rm s^{-1}}$ for $\mathfrak{g}=1$. Note that depending on the particular nature of the nonstandard theory of gravity, it may be the case that $\mathfrak{g}$ could take on different values for the Lense-Thirring and de Sitter effects \cite{Sab94,Ni10}.  

Lense-Thirring and de Sitter precession of classical angular momentum have been measured by satellite experiments. Gravity Probe B (GP-B), a satellite containing four highly spherical niobium-coated fused quartz gyroscopes in a cryogenic environment, measured the de Sitter precession of the rotational angular momentum of the gyroscopes to a 0.3\% precision and the Lense-Thirring precession of the gyroscopes to 20\% \cite{GravProbeB,Eve15}. A different approach was to use the satellite laser-ranging network \cite{laser2002} to precisely track the precession of the angular momentum associated with the orbital motion of the LAGEOS and LAGEOS II satellites themselves, rather than gyroscopes \cite{Ciu98}. Data from LAGEOS and LAGEOS II, combined with more recent data from the LAGOS satellite and a precise model of the Earth's gravitational field based on data from the GRACE satellite, measured the Lense-Thirring effect to a level of 0.5\% \cite{Ciu16,Lucchesi2019}.

\begin{table*}
\caption{Proposed characteristics of the orbiting ferromagnetic gyroscope (FG) system for a measurement of general-relativistic spin precession. The FG is assumed to be a fully magnetized cobalt sphere in vacuum with superconducting shielding as described in the text.}
\medskip \begin{tabular}{lll} \hline \hline
\rule{0ex}{3.0ex} Characteristic~~~~~~~~~~~~~~~~~~~~~~~~~~~~~~~~~~~~~~~~ & Notation~~~~~~~~~~~~~~~~~~~~~~~~~~~ & Approximate Value \\
\hline
\rule{0ex}{2.8ex} Radius & $r$ & 1~mm \\
\rule{0ex}{2.8ex} Mass density & $\rho$ & $8.86~{\rm g/cm^3}$ \\
\rule{0ex}{2.8ex} Mass & $M \approx 4\pi\rho r^3/3$ & $4 \times 10^{-2}~{\rm g}$ \\
\rule{0ex}{2.8ex} Moment of inertia & $I \approx 2Mr^2/5$ & $1.6 \times 10^{-4}~{\rm g \cdot cm^2}$ \\
\rule{0ex}{2.8ex} Number of polarized spins & $N$ & $4 \times 10^{20}$ \\
\rule{0ex}{2.8ex} Gilbert damping constant & $\alpha$ & 0.01 \\
\rule{0ex}{2.8ex} Ferromagnetic resonance frequency & $\omega_0$ & $10^{11}~{\rm s^{-1}}$ \\
\rule{0ex}{2.8ex} Gyroscopic threshold field & $B^* = N\hbar^2/(g\mu_BI)$ & $3 \times 10^{-10}~{\rm G}$ \\
\rule{0ex}{2.8ex} Gyroscopic threshold frequency & $\Omega^* = N\hbar/I$ & $3 \times 10^{-3}~{\rm s^{-1}}$ \\
\rule{0ex}{2.8ex} Operating magnetic field & $B$ & $10^{-11}~{\rm G}$ \\
\rule{0ex}{2.8ex} Larmor precession frequency & $\Omega_B$ & $10^{-4}~{\rm s^{-1}}$ \\
\rule{0ex}{2.8ex} Temperature & $T$ & $0.1~{\rm K}$ \\
\rule{0ex}{2.8ex} Background gas density & $n$ & $10^3~{\rm cm^{-3}}$ \\
\hline \hline
\end{tabular}
\label{Table:FG-characteristics}
\end{table*}

Our proposed experiment is modeled on GP-B, where the rotating niobium-coated fused quartz spheres are replaced by FGs. To evaluate the sensitivity, we assume that the FG is housed within a satellite similar to that used in the GP-B experiment \cite{GravProbeB} and referenced via a telescope to a remote star. For our sensitivity estimates, we assume an FG with characteristics as listed in Table~\ref{Table:FG-characteristics}: a spherical cobalt ferromagnet of radius $r \approx 1~{\rm mm}$ with remanent magnetization along $\hat{\mb{n}}$. The direction of the magnetic moment of the FG can be measured using a Superconducting QUantum Interference Device (SQUID) to detect the magnetic flux through a pick-up loop. A pick-up loop placed at a distance $d \approx 1~{\rm mm}$ away from the tip of the ferromagnet with loop radius $d \sin \theta_m \approx 0.8~{\rm mm}$, where $\theta_m \approx 54.74^\circ$ is the magic angle, maximizes the flux capture and would measure a changing magnetic flux of amplitude $\Phi \approx 100~{\rm G \cdot cm^2}$ as the FG precesses. The sensitivity of a low-temperature SQUID to flux change is $\delta \Phi \lesssim 10^{-13}~{\rm G \cdot cm^2 / \sqrt{ Hz }}$ \cite{Aws88,Use11,Hub08,Cla04v1}, which gives a detector-limited angular resolution for the FG of $\delta \theta\ts{det} \approx \delta \Phi / \Phi \lesssim 10^{-15}~{\rm rad}/\sqrt{\rm Hz}$. This translates to a detection-limited spin-precession resolution:
\begin{align}
\Delta \Omega\ts{det} \approx 10^{-15} \prn{t[{\rm s}]}^{-3/2}~{\rm s^{-1}}~.
\label{Eq:SQUID-limit}
\end{align}
Estimates show that the fundamental quantum noise limit for an FG is far below $\Delta \Omega\ts{det}$ \cite{Kim16}.

We estimate that the dominant source of statistical uncertainty in a satellite experiment using an FG to measure GR effects is not from the detector noise of the SQUID but rather from background gas collisions that impart angular momentum to the FG, causing random walk of its spin $\abrk{\mb{S}}$. Based on analysis of Ref.~\cite{Kim16}, and accounting for the spherical geometry of the FG, we find that the spin-precession resolution is limited to
\begin{align}
\Delta \Omega\ts{gas} \approx \frac{m r^2}{6 N \hbar} \sqrt{ \frac{n v\ts{th}^3}{ \pi t} }~,
\label{Eq:gas-limit}
\end{align}
where $m$ is the mass of the background gas (assumed to be He in our case since the system is under cryogenic conditions), $v\ts{th}$ is the average thermal velocity of the background gas, and other relevant parameters are listed in Table~\ref{Table:FG-characteristics}, assuming a background-gas density corresponding to cryogenic ultrahigh vacuum \cite{Han08}. The effects of other sources of noise are estimated to be negligible compared to the effects of background gas (see Ref.~\cite{Kim16} and the Supplemental Material).

Using a ferromagnet as a gyroscope requires exquisite shielding and control of magnetic fields in order to avoid systematic errors due to magnetic torques. We propose to use a multi-layer superconducting Pb shielding system based on the GP-B design as described in Refs.~\cite{Tab94,Mes00} combined with a conventional multi-layer $\mu$-metal shielding and magnetic-field-control coil system as described, for example, in Ref.~\cite{Yas13}. To achieve ultralow magnetic fields, the $\mu$-metal/coil system, with feedback provided by internal SQUID magnetometers, is used to achieve an ambient magnetic field less than $10^{-11}~{\rm G}$, close to the noise limit of SQUID magnetometers for integration times of one second. Nested collapsed Pb foil shields are inserted within the $\mu$-metal/coil system and subsequently cooled below the superconducting phase transition. The collapsed Pb foil shields are folded in such a manner as to minimize their internal volume. Once the temperature of the Pb is below the superconducting phase transition, the shields are expanded by unfolding them so that they have a considerably larger internal volume. Persistent currents in the superconducting shields keep the flux constant and thus the field within the expanded Pb shields is reduced by the ratio between the effective areas of the expanded and collapsed Pb foil shield. In practice, the residual field can be reduced by a factor of more than a hundred per layer, with practical limitations due to thermoelectric currents generated in the Pb shield. For such a superconducting shield system, the magnetic field within the shield will be determined by the frozen flux. These techniques can be used to achieve a magnetic field at the position of the FG much smaller than the required threshold field for operation ($B^* \approx 3 \times 10^{-10}~{\rm G}$, see Table~\ref{Table:FG-characteristics} and Ref.~\cite{Kim16}).

The proposed size and geometry for the FG (a mm-diameter sphere) is motivated by the need to minimize perturbations from background gas collisions ($\Delta \Omega\ts{gas} \propto 1/N$ and minimized for a spherical shape), achieve the best possible detector-limited sensitivity ($\Delta \Omega\ts{det} \propto 1/N$ \cite{Kim16}), and maintain a reasonable requirement for the threshold field $B^*$.

Undoubtedly, some residual magnetic field $\v{B}$ within the shields will persist, and so the questions now become whether the FG precession frequency $\v{\Omega}_B$ due to this field is sufficiently stable and whether $\v{\Omega}_B$ can be reliably distinguished from the sought-after effects, $\v{\Omega}\ts{LT}$ and $\v{\Omega}\ts{dS}$. Superconductors can achieve remarkable stability: drifts at the level of a part in $10^{11}$ per hour have been measured \cite{Van99}. Assuming the residual trapped field in which the FG operates is $B \sim 10^{-11}~{\rm G}$, this leads to a magnetic field drift of $\approx 3 \times 10^{-26}~{\rm G/s}$, which corresponds to a drift of the magnetic precession frequency of $d\Omega_B/dt \approx 2 \times 10^{-19}~{\rm s^{-2}}$. For the purposes of these estimates, we assume the worst-case scenario of a linear magnetic field drift at this rate (although on long time scales the drift will likely be a random walk of $\v{B}$ and $\v{\Omega}_B$).

The stability of $\v{\Omega}_B$ is crucial for distinguishing magnetic precession from the Lense-Thirring and de Sitter effects.  For a residual field with $B \sim 10^{-11}~{\rm G}$, $\Omega_B \approx 10^{-4}~{\rm s^{-1}}$, which is much larger than the Lense-Thirring and de Sitter effects [Eqs.~\eqref{eq:LT} and \eqref{eq:dS}], and thus it is important to find a way to distinguish $\v{\Omega}_B$ from $\v{\Omega}\ts{LT}$ and $\v{\Omega}\ts{dS}$. In the case of the Lense-Thirring effect, $\v{\Omega}\ts{LT}$ periodically varies in time in a predictable way because $\hat{\v{R}}$ changes in time with respect to $\v{\Omega}_E$ as the FG orbits the Earth. If the FG is placed in an elliptical orbit, both $\v{\Omega}\ts{LT}$ and $\v{\Omega}\ts{dS}$ could be modulated by order unity as $R$ changes. Thus it would become possible to search for the predictable periodic variation of $\v{\Omega}\ts{LT}$ and $\v{\Omega}\ts{dS}$ on top of the stable background magnetic-field precession. An example of how this can be done is discussed in the Supplemental Material.

Further discrimination of $\v{\Omega}\ts{LT}$ and $\v{\Omega}\ts{dS}$ from $\v{\Omega}_B$ can be obtained by using an array of FGs and taking advantage of the vectorial nature of the general-relativistic spin-precession. Consider, for example, the Lense-Thirring effect (similar arguments can be made for the de Sitter effect). If $\v{\Omega}_B$ is parallel with $\v{\Omega}\ts{LT}$, the effects add linearly to the measured spin-precession frequency: $\Omega \approx  \Omega_B + \Omega\ts{LT}$. However, if $\v{\Omega}_B$ is perpendicular to $\v{\Omega}\ts{LT}$, the contribution of the Lense-Thirring effect is quadratically suppressed: $\Omega \approx  \Omega_B + \Omega\ts{LT}^2/\prn{ 2\Omega_B }$. An array of FGs in separate shields can be employed with magnetic fields oriented in different directions, such that the various FGs have different predictable periodic patterns of sensitivity to general-relativistic spin-precession effects. This will enable coherent averaging and suppress systematic errors due to field drift and local perturbations.

Additionally, it may be possible to rotate or modulate $\v{B}$ at a frequency much faster than the orbital frequency in order to further discriminate $\v{\Omega}\ts{LT}$ and $\v{\Omega}\ts{dS}$ from $\v{\Omega}_B$. This may be achieved by rotating the magnetic shielding relative to the FG since the residual magnetic field will be dominated by frozen flux rather than the finite shielding factor. Further mechanisms to improve signal detection are possible: if two types of ferromagnetic materials are used, such that the materials' gyromagnetic ratios are opposite, their magnetic precession is in opposite directions but the relativistic precession are in the same direction. For control of systematic errors, it may also be interesting to consider experiments with materials having high net spin polarization but negligible magnetization, high magnetization but negligible spin polarization, and varying ratios of quantum orbital angular momentum to intrinsic spin, such as used in torsion pendulum experiments measuring exotic spin-dependent interactions \cite{Rit93,Hou01,Hec13}.

\begin{figure}
\includegraphics[width=8.4cm]{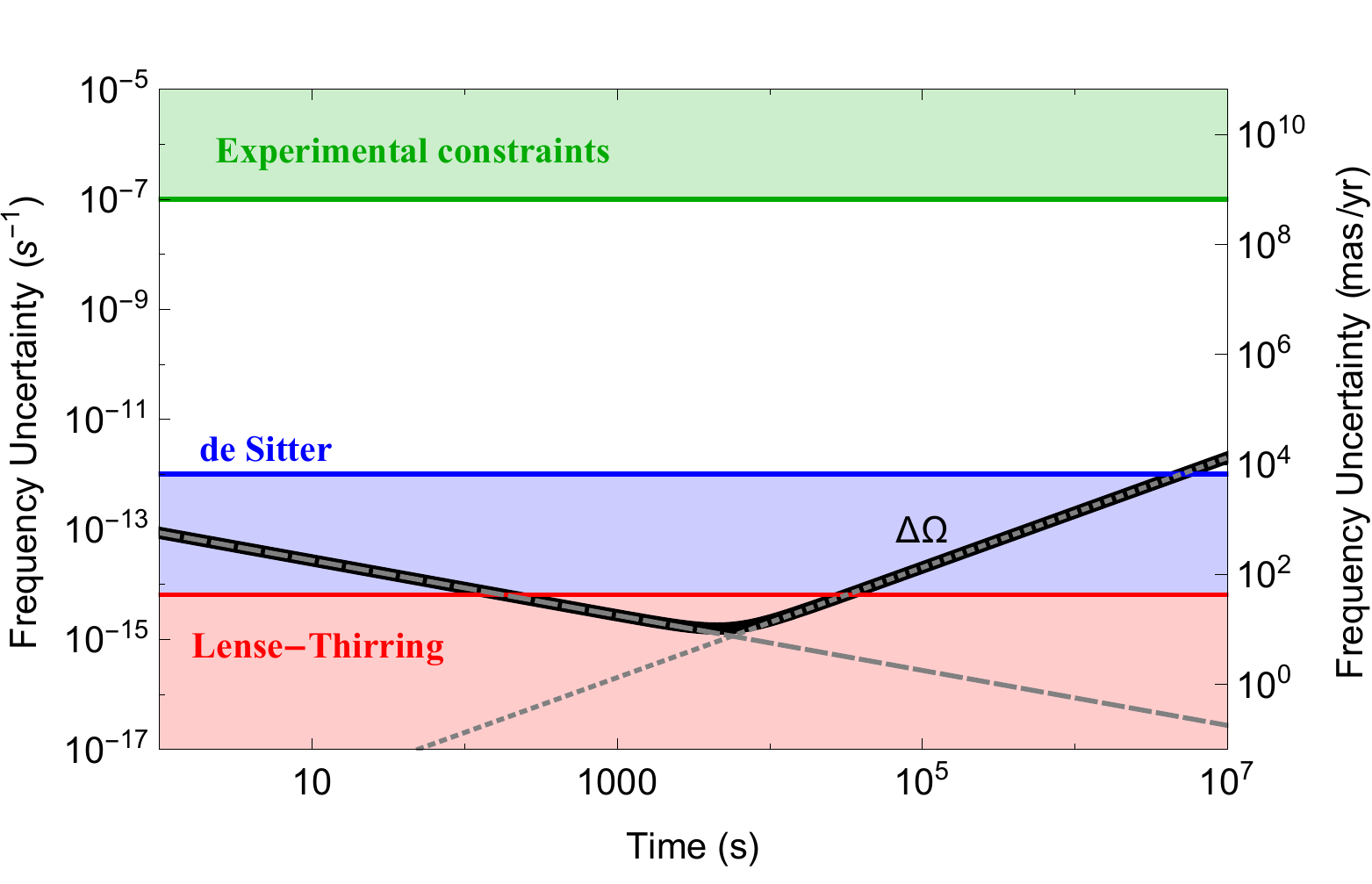}
\caption{Sensitivity to general relativistic spin-precession effects in the proposed ``Gravity Probe Spin'' experiment. The vertical scale on the right is in units of milliarcseconds (mas) per year. The black curve shows the projected uncertainty $\Delta\Omega$ in the measurement of the precession frequency $\Omega$ using a 1-mm radius spherical FG under conditions listed in Table~\ref{Table:FG-characteristics}. This curve results from two contributions summed in quadrature. First, the short-term statistical uncertainty is dominated by background gas collisions [Eq.~\eqref{Eq:gas-limit}, dashed gray line]. Second, The long-term uncertainty in the measurement is expected to be dominated by magnetic field drift within the superconducting magnetic shields, here assumed to be linear with rate $3 \times 10^{-26}~{\rm G/s}$ (dotted gray line). The blue line and light blue shaded area indicate the level beyond which the measurements are sensitive to the de Sitter effect \cite{Sitter1916,Schouten1918,Fokker1920} and the red line and pink shaded area indicate the level beyond which the measurements are sensitive to the Lense-Thirring effect \cite{LT1918,Schiff60}, calculated for the GP-B orbit and gyrogravitational ratio $\mathfrak{g}=1$ [Eqs.~\eqref{eq:LT} and \eqref{eq:dS}]. The green line and light green shaded area show existing experimental constraints on anomalous gravity-induced spin-precession \cite{Ven92,Hec08,Kim17}.}
\label{Fig:SensVsTime}
\end{figure}

Relative motion between the SQUID pick-up loop and the freely floating FG is another source of noise and systematic error that will require precise control. Errors due to this relative motion will ultimately be limited by the satellite position/orientation feedback control system referenced to the star-tracking telescope. We assume a star-tracking telescope and position/orientation feedback control similar to that used by GP-B, which had a long-term accuracy corresponding to $5\times 10^{-10}~{\rm rad}$ \cite{Dou95,Gwo03,Con15}, which would provide sufficient stability for measurement of the $\mathfrak{g}=1$ de Sitter and Lense-Thirring effects. Related technical issues are the trapping and release of the FG once the satellite is in orbit, damping of rotational motion of the FG such that $L \ll S$, vibrations of the pick-up coil, and the effect of electrostatic and magnetic forces on the FG that might accelerate the FG relative to the satellite housing. Protocols for measurement and control of the FG and pick-up coil motion will need to be designed and could, for example, involve damping of FG motion using eddy currents \cite{Tao2019} induced in a retractable conductor or various trapping and cooling techniques that have been developed to control the motion of macroscopic objects \cite{Teu11,Jas11}. The effects of stray electric fields and patch potentials, important issues for GP-B \cite{GravProbeB}, are considered in the Supplemental Material. Considering all such sources of noise and systematic errors, we expect that the ultimate accuracy of an FG-based measurement of general relativistic spin precession will be determined by the SQUID sensitivity, collisions of residual gas molecules with the FG, and magnetic field drift.

Figure~\ref{Fig:SensVsTime} shows the scaling of uncertainty in the measurement of the spin precession frequency $\Omega$ as a function of time considering the aforementioned effects. In principle, the projected measurement sensitivity of such a ``Gravity Probe Spin'' experiment is sufficient to measure the de Sitter and Lense-Thirring effects for $\mathfrak{g}=1$.  Consequently, stringent bounds will result on parametrized post-Newtonian (PPN) physics, scalar-tensor theories, and other standard-model extensions \cite{Overduin2015}. By comparing the sensitivity of Gravity Probe Spin to existing experimental bounds on anomalous gravity-induced spin-precession \cite{Ven92,Hec08,Kim17} as shown in Fig.~\ref{Fig:SensVsTime}, the proposed experiment has the potential to explore many decades of unconstrained parameter space.

In conclusion, we have described a satellite experiment using mm-scale ferromagnetic gyroscopes that has the potential to perform the first measurement of gravitational frame-dragging of intrinsic spins of electrons. This experiment, building on the technology of Gravity Probe B, would be a unique test at the intersection of quantum mechanics and general relativity. While such an experimental program requires extensive further studies of possible sources of noise and systematic errors, we hope that the long-term possibility of such a test will further motivate ongoing experimental efforts to develop levitating ferromagnetic gyroscopes.

This research was supported by the Heising-Simons and Simons Foundations, the U.S. National Science Foundation under Grant No. PHY-1707875, the DFG through the DIP program (FO703/2-1), and by a Fundamental Physics Innovation Award from the Gordon and Betty Moore Foundation. The work of DB supported in part by the DFG Project ID 390831469: EXC 2118 (PRISMA+ Cluster of Excellence), the European Research Council (ERC) under the European Union Horizon 2020 Research and Innovation Program (grant agreement No. 695405), and the DFG Reinhart Koselleck Project.
The work of AS supported in part by the US National Science Foundation grant 1806557, US Department of Energy grant DE-SC0019450, the Heising-Simons Foundation grant 2015-039, the Simons Foundation grant 641332, and the Alfred P. Sloan foundation grant FG-2016-6728.

\section{Supplemental Material}

\subsection{Magnetic torque noise}
\label{App-magnetic-torque-fluctuations}

An additional source of error affecting an FG, not considered in Ref.~\cite{Kim16}, was pointed out in Ref.~\cite{Band2018} (see also Appendix~\ref{App-modeling}). As noted in Ref.~\cite{Kim16}, the spin-lattice coupling generates stochastic fluctuations of the FG's magnetic moment $\bs{\mu}$ described by the fluctuation-dissipation theorem. In the presence of a nonzero magnetic field $\mb{B}$, this leads to a stochastic $\bs{\mu}\times\mb{B}$ torque acting on the FG, which in turn causes a random walk of the FG's spin axis $\abrk{\mb{J}}$. This coupling of the FG to the external environment through $\mb{B}$ generates noise in a measurement of the precession frequency:
\begin{align}
\Delta \Omega\ts{B} \approx \frac{\Omega_B^2}{\omega_0\Omega^*} \sqrt{\frac{4\alpha k_B T}{\hbar N t}}~,
\label{Eq:field-limit}
\end{align}
where $k_B$ is Boltzmann's constant. Under the conditions of our proposed experiment, $\Delta \Omega\ts{B}$ is significantly smaller than other sources of error.

\subsection{Electric field requirements}
\label{App-patch-potentials}

A precessing FG located in a spatial region with non-vanishing electric field may experience an electric-field-induced torque. In this section we estimate the requirements on the electric field and its gradient, in order to keep the FG precession rate due to this torque below the expected signal level.

A conducting sphere in a uniform electric field experiences no torque, since the induced electric dipole moment is parallel to the electric field. However a slight deviation from a spherical shape breaks the symmetry of the polarizability tensor, and, in general, causes the induced dipole moment to be at an angle to the electric field. Assuming the FG is shaped as a prolate spheroid (with semi-axes $a$, $b$, and $c$, where $a>b=c$) with small eccentricity $\varepsilon=\sqrt{1-b^2/a^2}$, the correction to the depolarization factors is of order $\varepsilon^2$~\cite{Landau8}. The torque on such a slightly non-spherical FG of radius $r$ in a uniform electric field $E$ can be estimated (in cgs units) as $\tau_e^{(1)}\approx\varepsilon^2r^3E^2/5$. The resulting precession rate is given by $\Omega_e^{(1)}=\tau_e^{(1)}/(N\hbar)$. The requirement to keep this rate below $\Omega\ts{LT}$ with $\mathfrak{g} = 1$, $\Omega_e^{(1)} \lesssim 4 \times 10^{-14}$~s$^{-1}$, imposes the following condition on the product between the eccentricity and the magnitude of the electric field:
\begin{align}  \label{e1}
\left| \varepsilon E \right| \lesssim 3 \times 10^{-6}\,{\rm V/cm}.
\end{align}
It should be noted that, in practice, the requirement on $|\varepsilon E|$ may be significantly reduced since orbital modulation can be used to distinguish general relativistic precession effects from nominally constant background torques, as discussed in Sec.~\ref{App-FG-orbital-dyamics}.

An electric field gradient $E'$ will exert a force on the FG, which must balance with all the other forces in the FG at its equilibrium point. Since there are certainly other forces, there may be a non-vanishing electric field gradient, which exerts a torque on the FG even if it is a perfect sphere. The magnitude of this torque can be estimated as $\tau_e^{(2)}\approx r^4EE'$. The resulting precession rate is given by $\Omega_e^{(2)}=\tau_e^{(2)}/(N\hbar)$. The requirement to keep this rate below $\Omega\ts{LT}$ with $\mathfrak{g} = 1$, $\Omega_e^{(2)} \lesssim 4 \times 10^{-14}$~s$^{-1}$, imposes the following condition on the product between the electric field and the gradient:
\begin{align}  \label{e2}
\left| EE' \right| < 10^{-11}\,{\rm V^2/cm^3}.
\end{align}
A procedure to reduce systematic error due to $\tau_e^{(2)}$, often employed in precision measurement protocols, is to apply a large electric field $E$ and use a measurement of $\Omega_e^{(2)}$ to minimize $E'$, then apply a large electric field gradient $E'$, and use a measurement of $\Omega_e^{(2)}$ to minimize $E$. Performed iteratively, this procedure can enable cancellation of residual $E$ and $E'$ to relatively high precision, and will also help reduce systematic error due to nonsphericity of the FG [Eq.~\eqref{e1}]. 

The electric field at the equilibrium position of the FG is created by potentials on proximal surfaces. To control electric fields these surfaces have to be coated with a high-conductivity material, such as gold. Nonetheless, surface-potential patches of order 10~mV are still likely to be present~\cite{Kim2010}. The electric field from such patches falls off exponentially with distance to the surface. We estimate that 10~mV patches with spatial scale of $<1$~mm create electric fields that satisfy requirements described by Eqs.~\eqref{e1} and \eqref{e2} provided the FG is $>1$~cm away from the surface. These estimates give the requirements on the  surface preparation necessary to ensure that electrostatic precession remains below the GR signal. Again, FG precession due to GR effects can be distinguished from $\Omega_e^{(1)}$ and $\Omega_e^{(2)}$ through orbital modulation as described in Sec.~\ref{App-FG-orbital-dyamics}.

\subsection{Model of ferromagnetic gyroscope dynamics}
\label{App-modeling}

We model the FG dynamics using the formulation described in Ref.~\cite{Band2018}. The FG is taken to be a single-domain spherical magnet with body-fixed moments of inertia ${\cal I}_X = {\cal I}_Y = {\cal I}_Z \equiv {\cal I}$. It is subject to a uniform magnetic field ${\bf B}$ and general-relativistic precession described by the angular velocity vector ${\boldsymbol \Omega}_r$. The Hamiltonian describing this system is given by:
\begin{align}  \label{Ham_top_E_field}
\hat{H} = \underbrace{\frac{1}{2 {\cal I}} {\hat {\bf L}}^2}_{H_R}  \underbrace{-(\omega_0/\hbar)  ({\hat {\bf S}} \cdot \hat{{\bf n}})^2}_{H_A}
  \underbrace{ - \hat{{\boldsymbol \mu}} \cdot {\bf B}}_{H_B} + \underbrace{ {\boldsymbol \Omega}_r \cdot ({\hat {\bf L}} + \mathfrak{g}{\hat {\bf S}}) }_{H_\Omega} \, .
\end{align}
In the rotational Hamiltonian $H_R$, ${\hat {\bf L}}$ is the orbital angular momentum operator; in the anisotropy Hamiltonian $H_A$ \cite{Brown_63}, ${\hat {\bf S}}$ is the spin operator, ${\hat {\bf n}}$ is the operator for the unit vector in the direction of the easy magnetization axis, and $\omega_0$ is the ferromagnetic resonance frequency; in the Zeeman Hamiltonian term $H_B$, $\hat{{\boldsymbol \mu}} = g \mu_B {\hat {\bf S}}$ is the magnetic moment operator ($\mu_B$ is the Bohr magneton and $g$ is the Land\'e factor); and ${H_\Omega}$ is the Hamiltonian accounting for the angular velocity vector ${\boldsymbol \Omega}_r$ related to general-relativistic precession, where $\mathfrak{g}$ is the gyrogravitational ratio (if $\mathfrak{g}=1$ the GR effects for intrinsic spin $\bf{S}$ and orbital angular momentum $\bf{L}$ are the same).

The dynamics are treated semiclassically since the FG has large spin expectation value $\abrk{S}$, as done in Ref.~\cite{Band2018}.
We write the Heisenberg equations of motion in reduced units, defining dimensionless vectors: the
unit spin ${\bf m} \equiv {\bf S}/S$, the orbital angular momentum  ${\boldsymbol \ell} \equiv {\boldsymbol L}/S$, the
total angular momentum, ${\bf j} = {\bf m} + {\boldsymbol \ell}$ and the unit vector in the direction of the magnetic field
${\bf b}={\bf B}/B$:
\begin{align} \label{dot_S_simple_5}
    {\dot {\bf m}} &= \omega_B {\bf m} \times {\bf b} + \omega_0 ({\bf m} \times {\bf n}) ({\bf m} \cdot {\bf n})
     \nonumber \\
    &~~~- \alpha {\bf m} \times ({\dot {\bf m}} - {\boldsymbol \Omega} \times {\bf m}) + \mathfrak{g}\prn{{\boldsymbol \Omega}_r \times {\bf m}} \, ,
\\ \nonumber \\ \label{dot_L_simple_5}
    {\dot {\boldsymbol \ell}} &= - \omega_0 ({\bf m} \times {\bf n}) ({\bf m} \cdot {\bf n})  \nonumber \\
    &~~~+ \alpha {\bf m} \times ({\dot {\bf m}} - {\boldsymbol \Omega} \times {\bf m}) + {\boldsymbol \Omega}_r \times {\boldsymbol \ell} \, ,
\\ \nonumber \\ \label{dot_n_simple_5}
    {\dot {\bf n}} &= ({\boldsymbol \Omega} + {\boldsymbol \Omega}_r) \times {\bf n} \, ,
\end{align}
where the angular velocity vector ${\boldsymbol \Omega}$ is given by
\begin{eqnarray} \label{Eq:Omega}
{\boldsymbol \Omega} = \omega_1 {\boldsymbol \ell} = \omega_1 ({\bf j} - {\bf m}) \, .
\end{eqnarray}
Here $\omega_B = g \mu_B |{\bf B}|$ is the Larmor frequency and $\omega_1 = S/{\cal I}$ is the nutation frequency.  The terms containing the Gilbert damping coefficient $\alpha$ account for Gilbert dissipation of spin components perpendicular to the easy magnetization axis. The Gilbert damping is due to interactions of the spin with internal degrees of freedom such as lattice vibrations (phonons), spin waves (magnons), thermal electric currents, etc.~\cite{Callen_Welton,Gilbert_04}. The Gilbert damping tends to lock the spin to the easy axis because the components of the spin orthogonal to the easy axis quickly decay \cite{Band2018}.  Hence we take ${\bf m}(t) = {\bf n}(t)$, which also simplifies the numerical calculations. Adding the spin and rotational angular-momentum in Eqs. \eqref{dot_S_simple_5} and \eqref{dot_L_simple_5}, we obtain
\begin{align}
    \label{dot_j_simple_5}
    {\dot {\bf j}} &= {\dot {\bf m}} + {\dot {\boldsymbol \ell}} = \omega_B \prn{ {\bf m} \times {\bf b} } + {\boldsymbol \Omega}_r \times \prn{{\boldsymbol \ell}+\mathfrak{g}{\bf m}} \, ,
    \\ \nonumber \\
    \label{Eq:Dot_j_simple_5}
     &= \omega_B \prn{ {\bf m} \times {\bf b} } + {\boldsymbol \Omega}_r \times \sbrk{ {\bf j} + \prn{ \mathfrak{g} - 1 } {\bf m} } \, .
\end{align}
Using Eq.~\eqref{Eq:Omega} and our approximation that ${\bf m} = {\bf n}$ (hence ${\bf m}\times{\bf n} = 0$), Eq.~(\ref{dot_n_simple_5}) can be rewritten in the form
\begin{eqnarray} \label{Eq:Dot_n_simple_5}
    {\dot {\bf m}} &=& (\omega_1 {\bf j} + {\boldsymbol \Omega}_r) \times {\bf m} \, .
\end{eqnarray}

We can solve Eqs.~(\ref{Eq:Dot_j_simple_5}) and (\ref{Eq:Dot_n_simple_5}) for a given satellite trajectory that specifies ${\boldsymbol \Omega}_r(t) = \v{\Omega}^{(1)}_{\text{LT}}(t) + \v{\Omega}^{(1)}_{\text{dS}}(t)$ [see Eqs.~(\ref{eq:LT}) and (\ref{eq:dS})] to obtain the dynamics of the FG. The upper index $(1)$ in the expression for ${\boldsymbol \Omega}_r(t)$ sets $\mathfrak{g}=1$ in Eqs.~(\ref{eq:LT}) and (\ref{eq:dS}), since in the modelling $\mathfrak{g}$ is present in the dynamical equations such that it distinguishes between the effect of general-relativistic precession of intrinsic spin as compared to that of angular momentum, as seen in Eq.~\eqref{Ham_top_E_field}. The results of the modeling for illustrative cases are discussed in the next section.

\subsection{Orbital dynamics of ferromagnetic gyroscopes}
\label{App-FG-orbital-dyamics}

In order to use an FG to measure GR-induced spin precession, it is crucial to have a distinct signature that can be differentiated from background effects. As noted in the main text, periodic motion of an FG at harmonics of the orbital frequency arise due to the modulation of $\v{\Omega}\ts{LT}$ and $\v{\Omega}\ts{dS}$ as the FG orbits the Earth. This offers a method to distinguish GR-induced spin precession from Larmor precession and nutation, whose frequencies are constant in time for fixed $\mb{B}$, as can be seen from the discussion in Sec.~\ref{App-modeling}.

\begin{figure}[!ht]
\includegraphics[width=8.4cm]{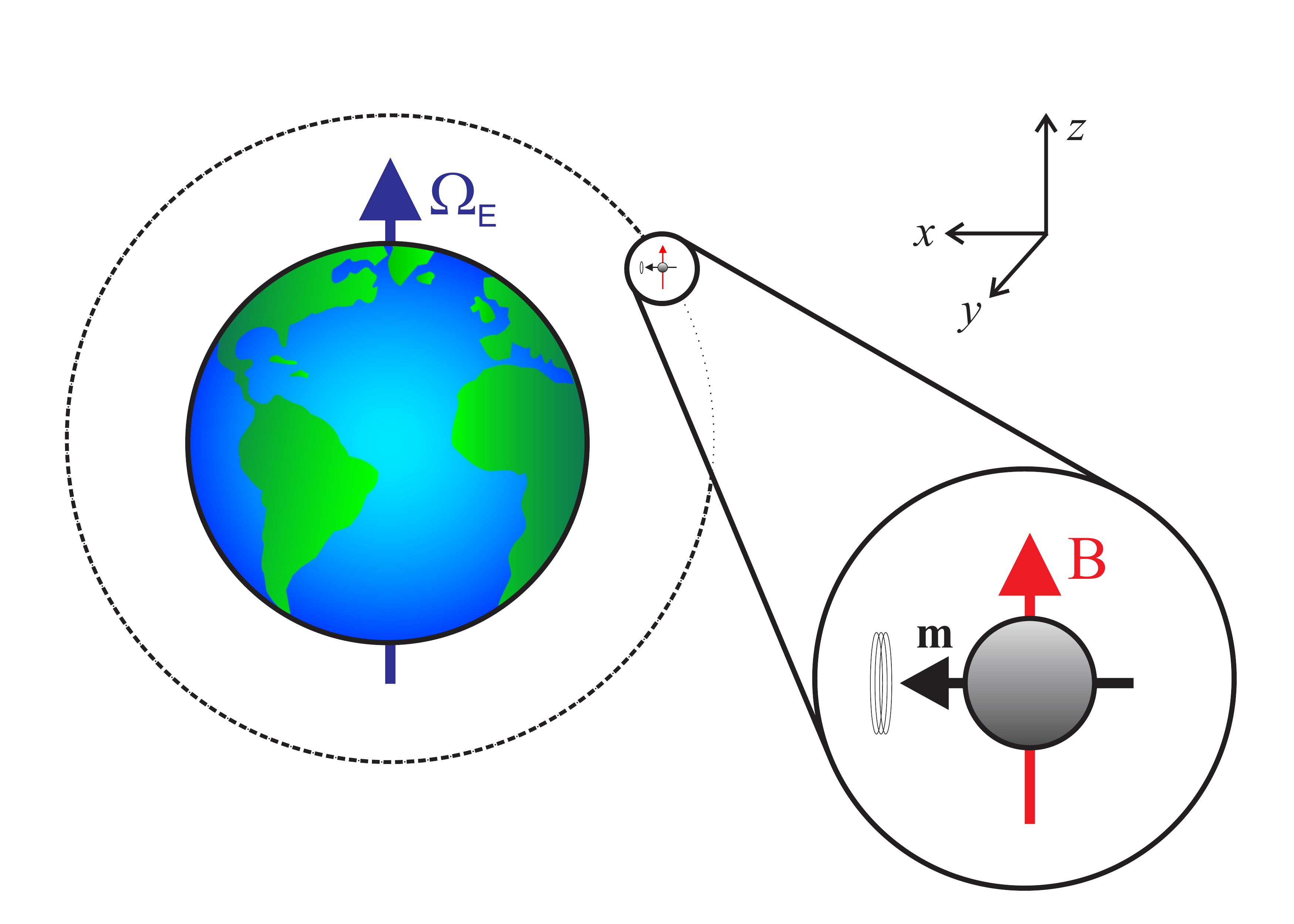}
\caption{Conceptual schematic diagram of a ``Gravity Probe Spin'' experiment. A freely floating spherical FG located within a superconducting shield is in a circular polar orbit. The magnetic field $\mb{B}$ (from the frozen flux in the superconducting shields) is oriented parallel to the direction of Earth's rotation axis $\bm{\Omega}_E$, both designated to point along $z$. The insert shows the initial orientation of the FG's magnetic moment and spin $\mb{m}$ along the $x$ axis. The pick-up coils measure the FG's magnetization along $x$. This geometry is designed for the detection of the Lense-Thirring effect.}
\label{Fig:InitialSetup}
\end{figure}

To illustrate the use of orbital modulation in a ``Gravity Probe Spin'' experiment, we model the behavior of an FG in a circular polar orbit around the Earth with radius $R \approx 7,000~{\rm km}$ (Fig.~\ref{Fig:InitialSetup}). The FG operates in an external magnetic field ${\bf B}$ oriented along Earth's rotation axis $\bm{\Omega}_E$, chosen to be the $z$-axis of our coordinate system. As discussed in Sec.~\ref{App-modeling}, the spin is locked along the direction of its easy magnetization axis by Gilbert damping, and is initially prepared to be perpendicular to $\mb{B}$, along $x$. In this geometry, precession due to the de Sitter effect [Eq.~\eqref{eq:dS}] is both constant in time, since $R$ is constant, and quadratically suppressed, since $\bm{\Omega}\ts{dS}$ is perpendicular to $\bm{\Omega}_B$ and $\Omega\ts{dS} \ll \Omega_B$. On the other hand, the Lense-Thirring precession $\bm{\Omega}\ts{LT}(t)$ is parallel to $\bm{\Omega}_B$ when the FG is at the north and south poles and is modulated at twice the orbital frequency [Eq.~\eqref{eq:LT}]. The orbital modulation of $\bm{\Omega}\ts{LT}(t)$ can be understood based on the fact that the Lense-Thirring effect generated by the rotation of the Earth is the gravito-magnetic equivalent of a dipole field, and possesses axial symmetry about $z$.
 
\begin{figure}[!ht]
\includegraphics[width=8.4cm]{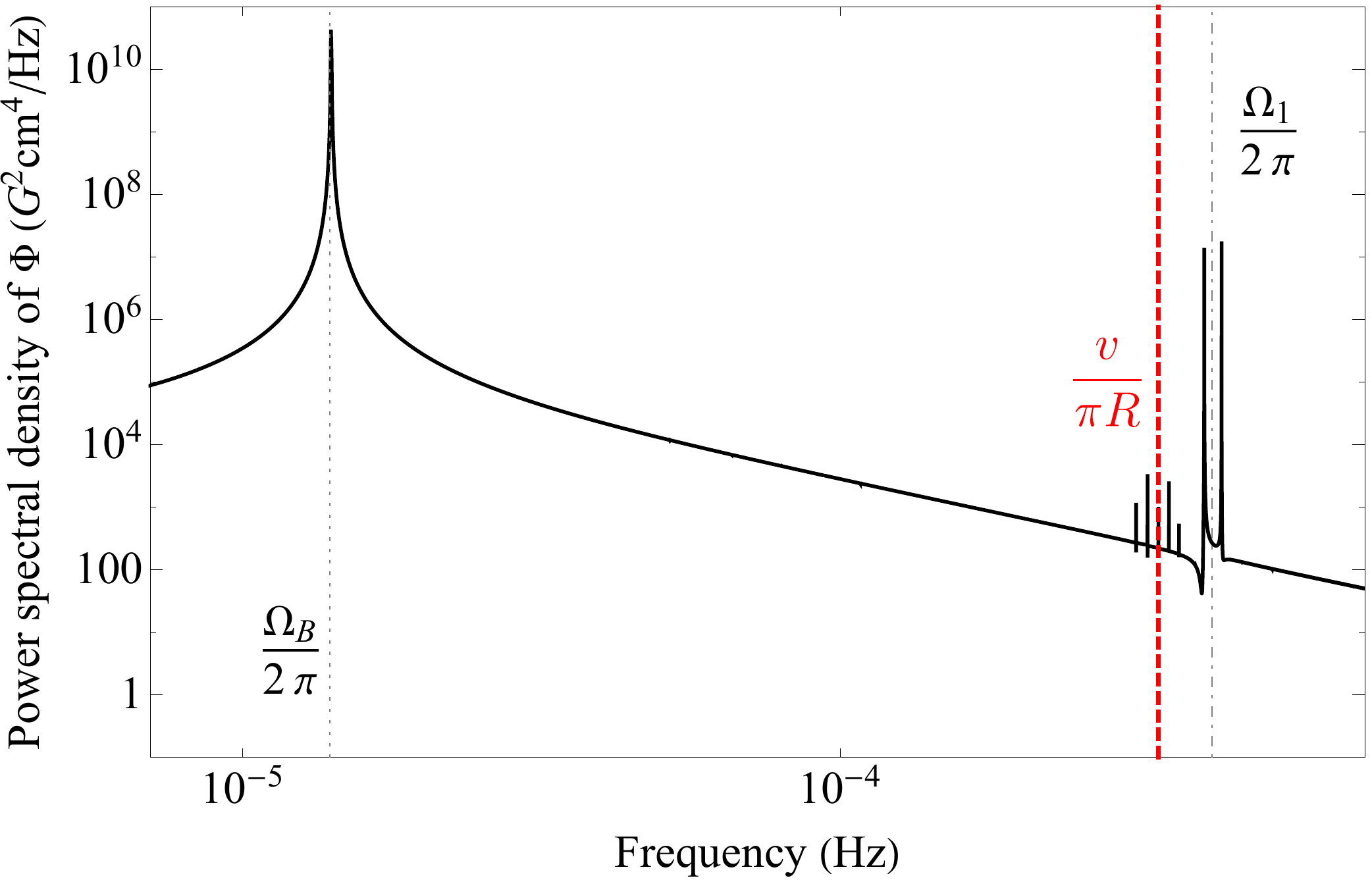}
\caption{Estimated power spectral density (PSD) of the time-dependent flux signal $\Phi$ due to a precessing FG that would be measured by a SQUID pick-up coil as in Fig.~\ref{Fig:InitialSetup}. The plot shows the PSD of a time-domain signal of duration $T = 3 \times 10^7$~s obtained by numerical solution of differential equations based on the model discussed in Sec.~\ref{App-modeling}. The parameters of the model match those listed in Table~\ref{Table:FG-characteristics}. The gray dotted line marks the Larmor frequency, $\Omega_B/(2\pi)$, the gray dot-dashed line marks the nutation frequency, $\Omega_1/(2\pi)$, and the red dashed line marks the second harmonic of the orbital frequency, $v/(\pi R)$. In order to enhance visualization, for this plot we choose $\mathfrak{g}=10^7$ for the Lense-Thirring effect, just below the present experimental constraints (Fig.~\ref{Fig:SensVsTime}).}
\label{Fig:GPS-PSD}
\end{figure}

\begin{figure}[!ht]
\includegraphics[width=8.4cm]{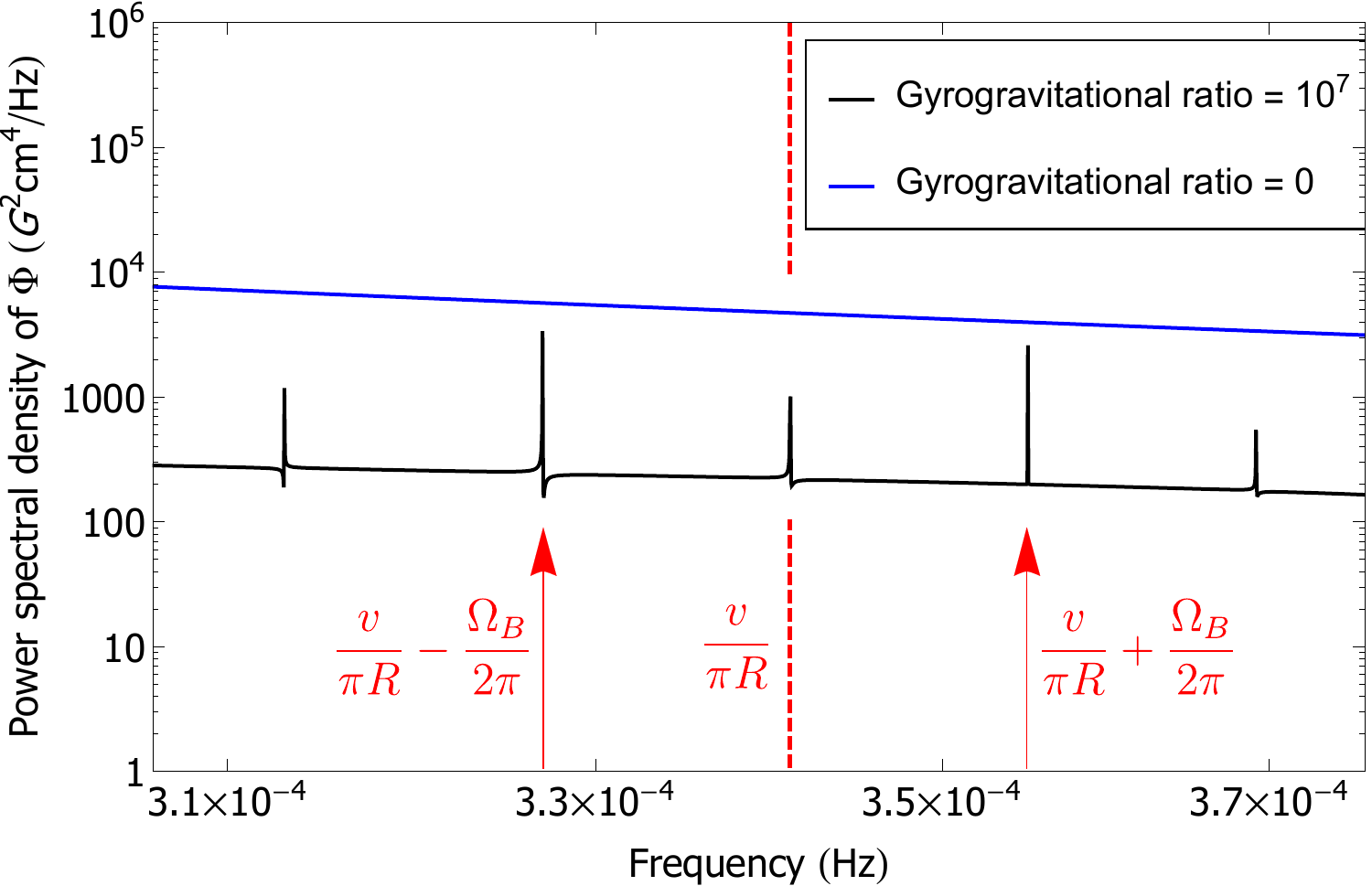}
\caption{The black curve shows the PSD of the time-dependent flux signal $\Phi$ under the same conditions and assumptions as in Fig.~\ref{Fig:GPS-PSD}. The blue curve, vertically offset for easier comparison, shows the PSD of the time-dependent flux signal $\Phi$ for the case where the gyrogravitational ratio $\mathfrak{g} = 0$. The dashed red line marks the second harmonic of the orbital frequency, $v/(\pi R)$, and prominent signals at sidebands shifted by the Larmor frequency are indicated by the red arrows at $v/(\pi R) \pm \Omega_B/(2\pi)$. Note also sidebands at $v/(\pi R) \pm \Omega_B/\pi$.}
\label{Fig:GPS-PSD-zoom}
\end{figure}

The results of a numerical solution of Eqs.~(\ref{Eq:Dot_j_simple_5}) and (\ref{Eq:Dot_n_simple_5}) for the FG dynamics, ${\bf m}(t)$, under the conditions described above are shown in Figs.~\ref{Fig:GPS-PSD} and \ref{Fig:GPS-PSD-zoom}. The figures show power spectral densities (PSDs) of the estimated flux $\Phi$ through a pick-up coil in the geometry described in the text [see discussion surrounding Eq.~\eqref{Eq:SQUID-limit}] as the FG orbits the Earth as shown in Fig.~\ref{Fig:InitialSetup}. In order to clearly discern the Lense-Thirring effect in Figs.~\ref{Fig:GPS-PSD} and \ref{Fig:GPS-PSD-zoom}, we choose $\mathfrak{g}=10^7$, just below the present experimental constraints on the Lense-Thirring effect (Fig.~\ref{Fig:SensVsTime}). 

\begin{figure}[!ht]
\includegraphics[width=8.4cm]{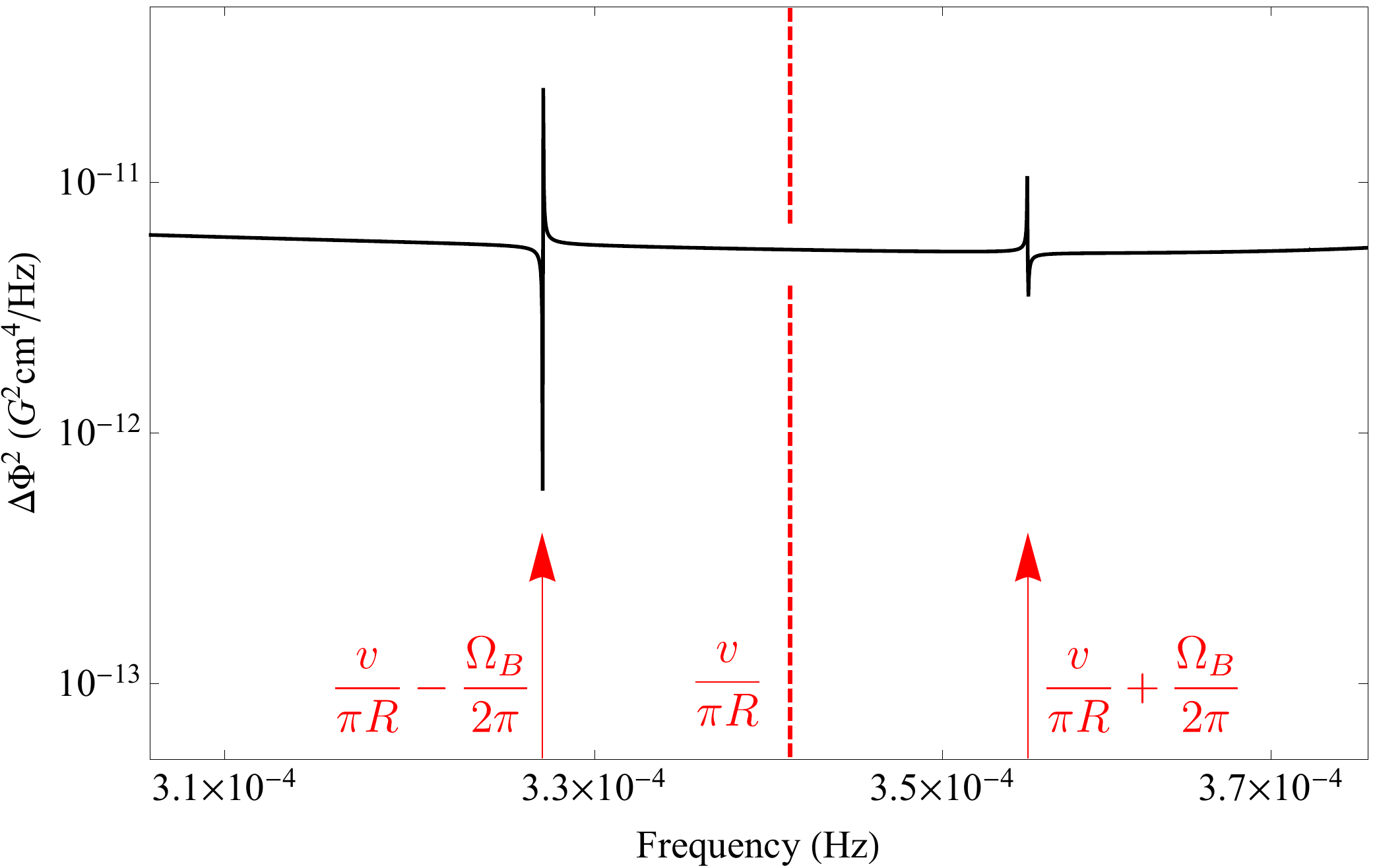}
\caption{PSD of the difference in time-dependent flux signal with $\mathfrak{g} = 1$ between two gyroscopes. The gyroscopes situated in opposite external magnetic fields along the $z$ axis. The conditions and assumptions are the same as in Fig.~\ref{Fig:GPS-PSD}. The dashed red line marks the second harmonic of the orbital frequency, $v/(\pi R)$, and prominent signals at sidebands shifted by the Larmor frequency are indicated by the red arrows at $v/(\pi R) \pm \Omega_B/(2\pi)$.}
\label{Fig:GPS-PSD-diff}
\end{figure}

The PSD shown in Fig.~\ref{Fig:GPS-PSD} demonstrates, as expected, that the dominant signal is at the Larmor frequency ($\Omega_B$) and prominent signals due to nutation appear at $\Omega_1$ with sidebands at $\Omega_1 \pm \Omega_B$. There is a noticeable signal due to the Lense-Thirring effect (with $\mathfrak{g}=10^7$) at the second harmonic of the orbital frequency, $2\pi \times v/(\pi R)$ (in rad/s, note the frequency units in the figures are Hz). In Fig.~\ref{Fig:GPS-PSD-zoom}, the signal with $\mathfrak{g}=10^7$ is compared to the signal for $\mathfrak{g}=0$ near the second harmonic of the orbital frequency, $2v/R$. Figure~\ref{Fig:GPS-PSD-diff} shows $\Delta\Phi^2$, the PSD of the difference between the measured flux from two FGs situated in magnetic fields with equal magnitudes but opposite directions ($\pm\hat{z}$) for the case where $\mathfrak{g} = 1$. The $\mathfrak{g}=1$ case would correspond to the case of particular interest where intrinsic spin and orbital angular momentum behave identically in general relativity. As in the case where $\mathfrak{g} = 10^7$, there are noticeable signals arising from modulation of FG precession at twice the orbital frequency due to the Lense-Thirring effect, seen at the sideband frequencies $2v/R \pm \Omega_B$. The results of the modeling demonstrate that the Lense-Thirring effect indeed modulates FG precession at the second harmonic of the orbital frequency, offering a signature of GR effects distinguishable from effects that do not vary periodically with the orbit. The asymmetric shapes of the peaks in Figs.~\ref{Fig:GPS-PSD}, \ref{Fig:GPS-PSD-zoom}, \ref{Fig:GPS-PSD-diff}, and subsequent plots are described by Fano line shapes \cite{Fano61} that result from the interference of the background and the resonances in the PSD.

For reference, the expected measurement noise floor due to collisions with residual background gas, based on Eq.~\eqref{Eq:gas-limit}, is estimated to be
\begin{align}
    \delta(\Phi^2)\ts{gas} \approx \frac{10^{-9}}{\sqrt{T}}~{\rm G^2 cm^4 / Hz}~.
\end{align}
Comparing $\delta(\Phi^2)\ts{gas}$ to the signals plotted in Fig.~\ref{Fig:GPS-PSD-diff} show that for a measurement times $T \gtrsim 10^4~{\rm s}$ the Lense-Thirring precession for $\mathfrak{g} = 1$ should be resolvable, consistent with the sensitivity estimates shown in Fig.~\ref{Fig:SensVsTime}.

\begin{figure}[!ht]
\includegraphics[width=8.4cm]{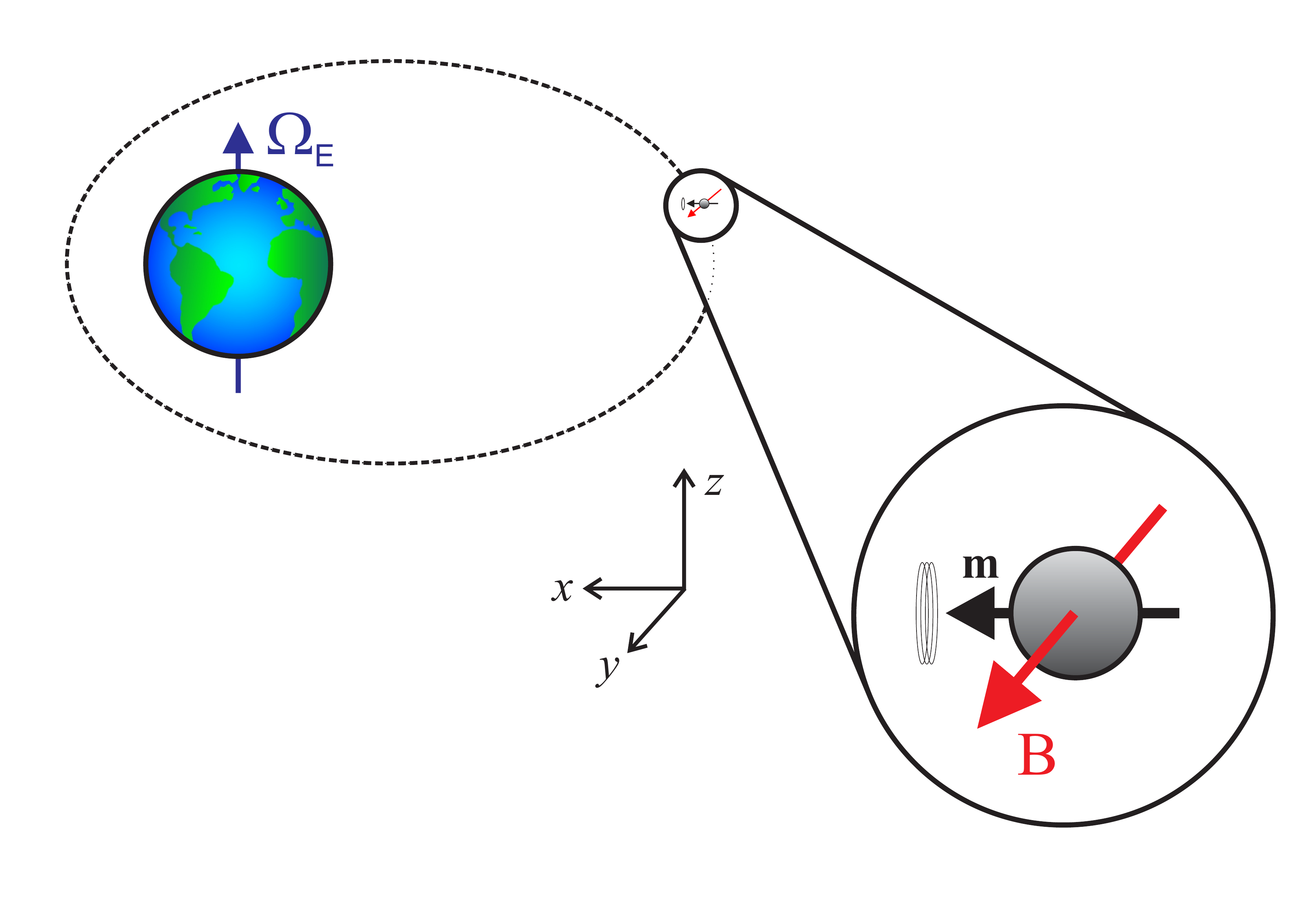}
\caption{Conceptual schematic diagram of a ``Gravity Probe Spin'' experiment similar to that shown in Fig.~\ref{Fig:InitialSetup} except that the orbit is elliptical and the magnetic field $\mb{B}$ is directed along the $y$-axis, perpendicular to the orbital plane. This geometry is designed for the detection of the de Sitter effect.}
\label{Fig:Setup-elliptical}
\end{figure}

\begin{figure}[!ht]
\includegraphics[width=8.4cm]{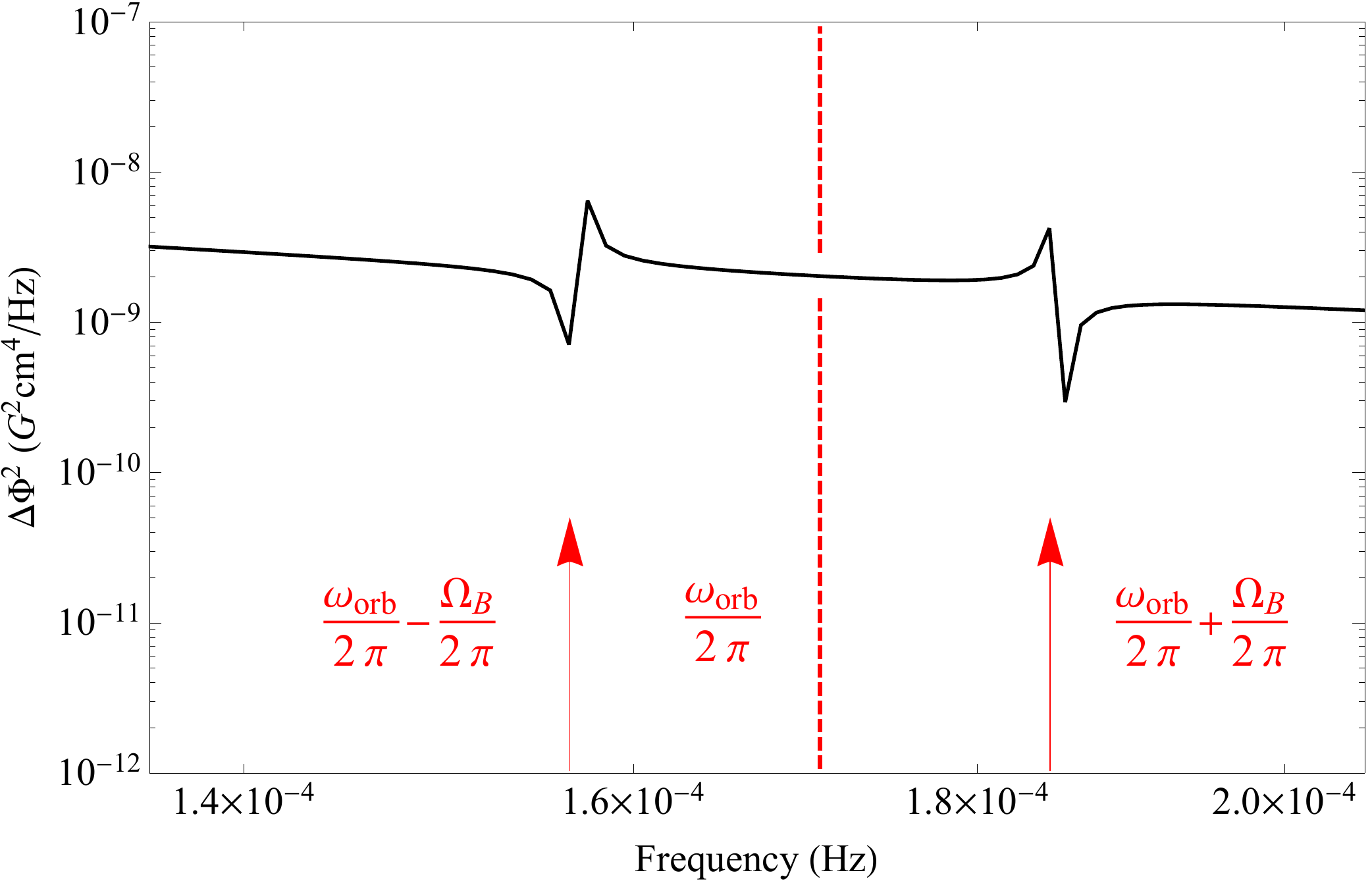}
\caption{PSD of the difference in time-dependent flux signal with $\mathfrak{g} = 1$ between two gyroscopes. The gyroscopes are situated, respectively, in external magnetic fields along the $y$ axis with equal magnitudes and opposite directions. The FG is modelled for the  duration of $10^6$~s in a polar elliptical orbit as indicated in Fig.~\ref{Fig:Setup-elliptical}, with ellipticity of 0.3. The dashed red line marks the first harmonic of the orbital frequency, $\omega\ts{orb}/(2\pi)$, and prominent signals at sidebands shifted by the Larmor frequency are indicated by the red arrows at $\omega\ts{orb}/(2\pi) \pm \Omega_B/(2\pi)$.}
\label{Fig:elliptical}
\end{figure}

Employing a different geometry for the FG, namely orienting $\mb{B}$ parallel to $\bm{\Omega}\ts{dS}$, gives linear sensitivity to $\bm{\Omega}\ts{dS}$ (in which case sensitivity to $\bm{\Omega}\ts{LT}$ is quadratically suppressed). By putting the satellite into an elliptical orbit (Fig.~\ref{Fig:Setup-elliptical}), $R$ and $\mb{v}$ are modulated and a distinct signature in the PSD of $\Phi$ can be obtained for the de Sitter effect, as demonstrated in Fig.~\ref{Fig:elliptical}. Figure~\ref{Fig:elliptical} gives the result of modeling the FG dynamics for a polar elliptical orbit with eccentricity of 0.3: the PSD shows the difference between the measured flux from two FGs situated in magnetic fields with equal magnitudes but opposite directions ($\pm\hat{y}$) assuming $\mathfrak{g} = 1$. Signals due to the de-Sitter effect are observed at sidebands around the orbital frequency $\omega\ts{orb}$, 
\begin{align}
\omega\ts{orb} = \sqrt{ \frac{G \, M}{a^3}} \, ,
\end{align}
where $a$ is the semi-major axis of the ellipse. This is expected since $\bm{\Omega}\ts{dS}(t)$ is periodic with the modulation of $R$ and $\mb{v}$ as the FG orbits, leading to a signal at the first harmonic of $\omega\ts{orb}$.

In conclusion, the numerical modeling demonstrates that, in principle, for particular experimental geometries there exist potentially measurable signatures of general relativistic precession of an FG at harmonics of the orbital frequency.

\end{document}